\documentclass[]{spie}  

\usepackage{array}
\usepackage{wrapfig}
\usepackage{multirow}
\usepackage{tabularx}
\usepackage{cite}
\usepackage{multicol}
\usepackage{balance}
\usepackage[colorlinks=true, allcolors=blue]{hyperref}
\usepackage[pdftex]{graphicx}
\usepackage{epstopdf}
\usepackage{float}

\usepackage{gensymb}
\usepackage{caption}
\usepackage{pgfplots,multicol}
\usepackage{amsmath,stmaryrd,pict2e,picture}
\usepackage{amsmath,amssymb}
\usepackage{mathtools}
\usepackage{pifont}

\newlength{\rowidth}
\AtBeginDocument{\setlength{\rowidth}{3em}}
\usepackage{mathrsfs}
\usepackage{algorithm, algorithmic}

\def\footnoterule{\relax%
	\kern-5pt
	\hbox to \columnwidth{\hfill\vrule width 0.75\columnwidth height 0.4pt\hfill}
	\kern4.6pt}
\makeatother
\title{Continuous Human Activity Recognition using a MIMO Radar for Transitional Motion Analysis}

\author{John Kobak}
\author{Bennett J. Richman}
\author{LaJuan Washington Jr.}
\author{Syed~A.~Hamza}

\affil{School of Engineering, Widener University, Chester, PA 19013, USA}

\authorinfo{E-mails: 	 \{jjkobak, bjrichman, lewashington, shamza\}@widener.edu
}

\begin{document}

\maketitle
\begin{abstract}
The prompt and accurate recognition of Continuous Human Activity (CHAR) is critical in identifying and responding to health events, particularly fall risk assessment. In this paper, we examine a multi-antenna radar system that can process radar data returns for multiple individuals in an indoor setting, enabling CHAR for multiple subjects. This requires combining spatial and temporal signal processing techniques through micro-Doppler (MD) analysis and high-resolution receive beamforming. We employ delay and sum beamforming to capture MD signatures at three different directions of observation. As MD images may contain multiple activities, we segment the three MD signatures using an STA/LTA algorithm. MD segmentation ensures that each MD segment represents a single human motion activity. Finally, the segmented MD image is resized and processed through a convolutional neural network (CNN) to classify motion against each MD segment.
\end{abstract}

\section{Introduction}

The use of contactless technology for detecting human motions has become increasingly popular due to its non-invasive nature as it eliminates the need for users to wear specific tracking devices \cite{7426551, 10.1117/12.2527660, 8894733}. Radar systems are particularly effective in this regard, as they provide reliable non-contact monitoring that is privacy-preserving and not affected by lighting conditions. Active  radio frequency (RF) sensing, in particular, allows for 4D imaging capabilities by measuring the scatterer's velocity in addition to range and 2D angular localization. This is a major advantage over visual-based systems, which require additional pre-processing and filtering operations to accurately detect small movements \cite{8010417, 8746862}. While a high-resolution imaging radar ensures privacy by generating silhouette-type portraits that reveal minimal identifiable information, unlike camera-based systems. Radar-based CHAR has diverse applications such as detecting significant events for automated surveillance, analyzing daily activities and behaviors, recognizing abnormalities in gait, monitoring health in care facilities and rehabilitation services, and promoting independent living for the elderly.  \cite{6126543, Amin2017RadarFI, 8610109, WANG2018118, 8613848, 9069251, 8746862}.

The primary objective of this research is to detect the real-time daily activities of one or multiple individuals residing in a shared dwelling. Utilizing radar technology to achieve effective CHAR presents a significant challenge, as it relies heavily on a series of pre-processing steps that must isolate both the individuals and their corresponding activities simultaneously. The proposed approach in this paper utilizes beamforming to filter the received data and achieve spatial filtering, effectively isolating the motion of a single individual in the field of view. As a result of human motion, the radar return undergoes a frequency shift due to the Doppler effect. However, since human movement involves the motion of various body parts, there are often additional movements and rotations in different parts of the body besides the torso's movement. For example, when a person walks, their arms swing naturally. These micro-scale movements result in additional Doppler shifts, known as micro-Doppler (MD) effects, which aid in identifying the motion type. After isolating an individual in a certain angular region, in the field of view, short-time Fourier transform is applied to the output of the beamformer to obtain a MD image. However, due to the limited angular resolution of the beamformer, the actions performed by individuals outside of the main beam may not be entirely separable, either due to their proximity to the other individual or leakage associated with the beamformer sidelobes. Hence, MD images might not entirely be representative of the activity of the individual in the main beam but also has remnants of the activities performed by other individuals at different locations. At this stage, it is possible that the MD image could include a series of activities, and the goal is to separate the series of activities executed by the individual present within the beamformer's main lobe. We apply STA/LTA (short-time-average/long-time-average trigger algorithm) in order to segment the MD image into sub-images, such that each sub-image comprise exclusively of a single motion. The STA/LTA algorithm performs event detection by utilizing two windows that operate on the envelopes of the MD image. The envelopes are extracted after smoothing out the modulated MD image, which facilitates the envelope detection algorithm employing the percentile technique. The STA/LTA algorithm's task of detecting the activity of a specific individual within the main beam is challenging due to the presence of residual activities from other individuals that were not intended to be captured in the MD image obtained after spatial filtering.  In the final step, the sub-images are padded to a uniform size and fed into a convolutional neural network (CNN) to classify and recognize the specific human activity being performed in the image sequence.  It is noted that padding is necessary for resizing the sub-images to achieve consistent input dimensions, as the sub-dimensions of images may vary depending on the activity. Two experimental studies were conducted to collect data, one involved a single subject while the other involved two subjects using beamforming. A total of 800 trials were performed, with various activities carried out by a maximum of two individuals.

The paper's organization is as follows:  Section 2  details the signal model discussing the methodology to process ADC data returns to convert them into visible MD spectrograms. Section 3 details the radar parameters and discusses the data collection process. Section 4 describes the procedure of cleaning the MD images and separating the individual events within a MD image. Section 5 outlines the structure of the proposed CNN for one person, and then extrapolate that to include a combination of right, left, and nominal spectrogram images. Finally, Section 6  reveals the results.
 

\section{Radar return signal analysis}
\label{Problem Formulation}
The complex valued raw  data matrix $\mathbf{s}(n,m) \in C^{N*M}$  of the frequency-modulated continuous
wave (FMCW) radar  is obtained through spatially processing the radar returns    by an  $M$ element uniformly spaced antenna array. The data is collected over  $N$ temporal sampling instances. The receiver array vector $\mathbf{s}(m) \in C^{M}$ at  time instant $n$ corresponds to the $n_{th}$ row of $\mathbf{s}(n,m)$ and is given by,
\begin {equation} \label{a}
\mathbf{s}(m)=    \sum_{l=1}^{L} \alpha _l \mathbf{a}( \theta_l)  + \mathbf{v}(m),
\end {equation}
where, $\mathbf{a} ({\theta_l})$  $\in \mathbb{C}^{M}$ is  the  steering vector   corresponding to the azimuth direction $\theta_l$ of the scatterer, and is defined  as follows,  
\vspace{+2mm}
\begin {equation}  \label{b}
\mathbf{a} ({\theta_l})=[1 \,  \,  \, e^{j (2 \pi / \lambda) d cos(\theta_l)  } \,  . \,   . \,  . \, e^{j (2 \pi / \lambda) d (M-1) cos(\theta_l)  }]^T.
\end {equation}
Here, $d$ is the inter-element spacing and $\alpha_l$ $\in \mathbb{C}$  is the complex amplitude of the radar return. The additive Gaussian noise $\mathbf{v}(m)$ $\in \mathbb{C}^M$ has   variance   $\sigma_v^2$.
The elements of the received data vector $\mathbf{s}(m)$ are combined linearly by the $M$-sensor beamformer that strives to spatially filter the reflections from all other directions except the signal in the direction of beamformer look angle $\theta_k$. The spatially filtered signal vector $\mathbf{x}({\theta_k})$ $\in \mathbb{C}^N$ after beamforming is given by, 
\begin {equation}  \label{c}
\mathbf{x}({\theta_k}) = \mathbf{s}(n,m) \mathbf{w}^H({\theta_k}),
\end {equation}
where $\mathbf{w}({\theta_k})=\mathbf{a}^H ({\theta_k})$ are the complex beamformer weights pointing towards $\theta_k$.

The spatially filtered signal vector $\mathbf{x}({\theta_k})$  is reshaped into a two-dimensional matrix, $\mathbf{x}_{\theta_k}(p, q)$. This is achieved by segmenting the $N$ dimensional vector $\mathbf{x}({\theta_k})$, such that, the $P$ samples collected within a pulse repetition interval (PRI) are stacked into a $P$ dimensional column. There are $Q$ such columns within $\mathbf{x}_\theta(p, q)$ where  $Q=N/P$  is the number of PRIs processed within the observation time $N$.  The range-map, $\mathbf{r}_{\theta_k}(p,q)$  is obtained by applying the column-wise Discrete Fourier Transform (DFT) operation which is given by,
\begin {equation}  \label{d}
\mathbf{r}_{\theta_k}(l,q) =  \sum_{p=0}^{P-1} \mathbf{x}_{\theta_k}(p, q)e^{-j (2 \pi l p/ N)}
\end {equation}
We observe the data in the time-frequency (TF) domain after localizing the motion in azimuth and range bins of interest. The spectrogram is used as the TF signal representation, showing the variation of the signal power as a function of time $n$ and frequency $k$. The spectrogram of a periodic version of
a discrete signal $\mathbf{v}_{\theta_k}(n)$, is given by \cite{30749, article4573,  9101078}, 
\begin {equation}  \label{e}
\mathbf{d}_{\theta_k}(n,k) = | \sum_{m=0}^{H-1} \mathbf{h}(m)\mathbf{v}_{\theta_k}(n-m)e^{-j (2 \pi k m/ H)}|^2,
\end {equation}
where $\mathbf{v}_{\theta_k}=\sum_{l=r_l}^{r_u}\mathbf{r}_{\theta_k}(l,q)$ is obtained by collapsing the range dimension beginning from lower range bin $r_l$ to highest range bin $r_h$. Tapering window $\mathbf{h}$ of length $H$  is applied to reduce the sidelobes. The spectrograms reveal the different velocities, accelerations, and higher order moments which cannot be easily modeled or assumed to follow specific non-stationary structures \cite{10.1117/12.669003, 295203}. We observe the motion of two persons performing different activities in close proximity to each other at different azimuth angles. We aim to correctly pair the activity to the corresponding azimuth angle. This is achieved by jointly processing the spectrograms, $\mathbf{v}_{\theta_1}(n,k)$ and $\mathbf{v}_{\theta_2}(n,k)$ which are respectively localized at azimuth angles $\theta_1$ and $\theta_2$. It is clear that the actions of multiple persons is hard to be distinguished in azimuth by only using a single antenna. 

\section{Data Collection}

To ensure the representation of the average human body type, we performed each transitional activity using six subjects with different heights, genders and body compositions. The number of trials conducted for each activity is outlined in Table \ref{table:1}. Please note that these classes arise after performing image segmentation on extended MD images. For single person activity, the extended MD images  were captured for 12s with different combinations of three human motions selected from five activities namely,  walking forward, walking backward, sitting, standing, and bending motions.  For the two-person data, trials were conducted for walking forward and walking back at $\pm30$\degree, resulting in four additional classes. The radar system used had a bandwidth of 4 GHz and operated at 77 GHz. The two participants performed activities while  facing the radar system with radial angles of {+30}\degree and {-30}\degree, respectively. The radar systems were positioned at an average distance of 2 meters from the participants, and the RF sensor's output transmission power was set to 40 mW.

\begin{itemize}
\item  PRI is set to 1 ms, and each data example is observed over the time period of 12 s, resulting in $Q=4000$ slow time samples.
\item  ADC sampling rate is 512 ksps, rendering 512 fast time samples per PRI, resultantly the length of data vector is $N=6156000$. 
\item  The  received data $\mathbf{s}(n,m) \in C^{N*M}$, is collected through $M=4$ element receive array, with an inter-element spacing of $\lambda/2$ ($\lambda$ is the wavelength corresponding to the operating frequency), therefore the dimensionality of received raw data matrix is $6156000\times4$. 
\item  Beamforming is performed on the raw data matrix, resulting in a spatially filtered $\mathbf{x}({\theta_k})$ vector of dimensions $3276800\times1$.  Two such vectors are generated in the directions of each motion ${\theta_1}$ and ${\theta_2}$.
\item  Each vector $\mathbf{x}({\theta_k})$
is reshaped into a $512\times12000$ matrix. After applying columnwise DFT,  and identifying the range bins of interest, the corresponding rows are summed together, resulting in  $\mathbf{v}_{\theta_k}=\sum_{l=r_l}^{r_u}\mathbf{r}_{\theta_k}(l,q)$, which is of dimensions $12000\times1$. 
\item  A combined spectrogram and two spectrograms after beamforming $\mathbf{d}_{\theta_1}$ and $\mathbf{d}_{\theta_2}$, each of dimensions $128\times128$ is obtained, where the window length is 128. 

\end{itemize}


\begin{table}[h]
\hbadness=99999
\centering

\vspace{1mm}

\begin{tabular}{ | m{3.8em} | m{3.5em} | m{3.8em}  | m{3.5em}  | m{3.5em} | m{3.5em}  | m{3.65em} | m{3.65em}  | m{3.8em}  | m{3.8em}  |} 
\hline
Examples & Forward & Backward & Bending & Standing & Sitting down & Forward +30\degree & Forward -30\degree & Backward +30\degree & Backward -30\degree  \\ 
\hline 
Total & 771 & 498 & 120 & 71 & 131 & 13 & 41 & 31 & 40\\
\hline
\end{tabular}

\caption{ Number of trials for segmented activity } 
\label{table:1}
\end{table} 

\hfill
\vspace{3mm}\textbf{}

\section{Exyended Micro-Doppler Segmentation}

\begin{figure}[H]
	\includegraphics[height=2.5in, width=6.5in]{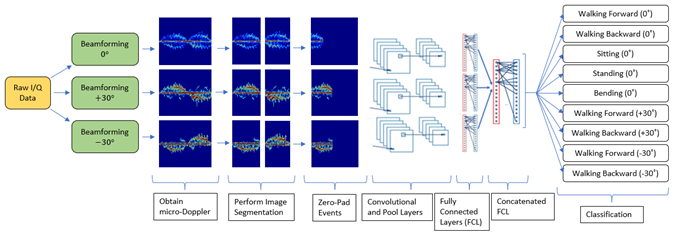}
    \caption{Flowchart for the proposed approach}
	\label{PreProcessing_Image.1}
\end{figure} 

Figure \ref{PreProcessing_Image.1} depicts the complete processing chain, beginning with radar data collection and culminating in the classification of a specific human activity type. Once the extended MD  images are acquired, after beamforming\cite{10.1117/12.2618718}, the envelope and event detection algorithms are applied to locate the start and end times of possible events within the duration of the extended MD image. These timestamps are then used to identify and crop the individual events within the extended MD  image. Next, the cropped images are resized to a resolution of $600\times600$ by zero-padding, enabling all of them to be consistent with the input dimensions of the CNN. Event detection is accomplished using the STA/LTA algorithm, which is elaborated on below.

\subsection{STA/LTA algorithm for Event Detection}

The short-time-average/long-time-average STA/LTA trigger algorithm is commonly utilized in weak-motion applications where the detection of events is most desired. With extended MD  returns providing TF characteristics of reflected Doppler signals, the STA/LTA algorithm can identify and extract the desired features by tracking the start and end times of events. The STA/LTA algorithm is applied to the envelope of the extended MD  which is obtained using an envelope detection algorithm. 

\begin{figure}[H]
    \center
    \includegraphics[height=1.8in, width=2.2in]{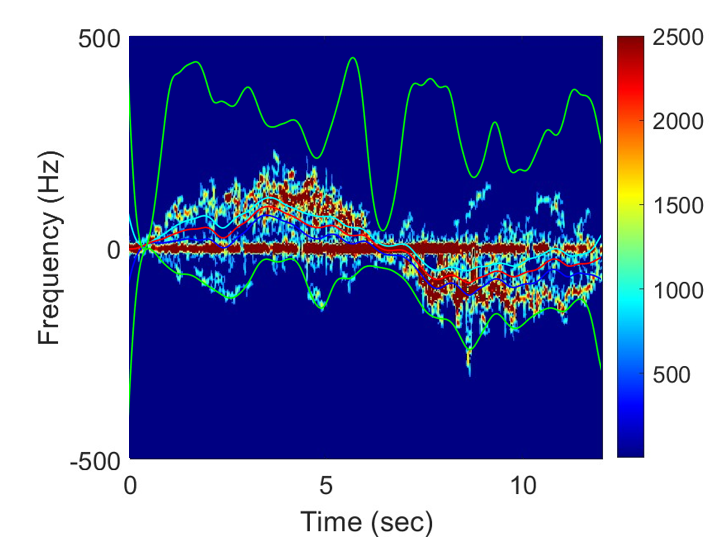}
    \includegraphics[height=1.8in, width=2.2in]{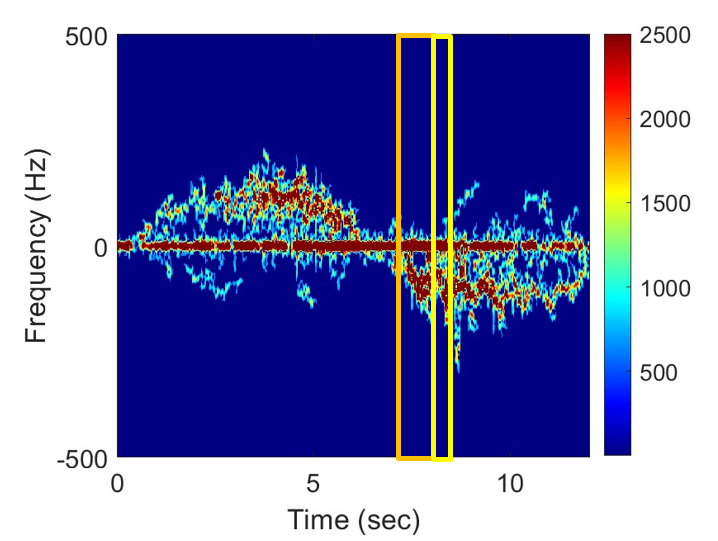}
    \includegraphics[height=1.8in, width=2.2in]{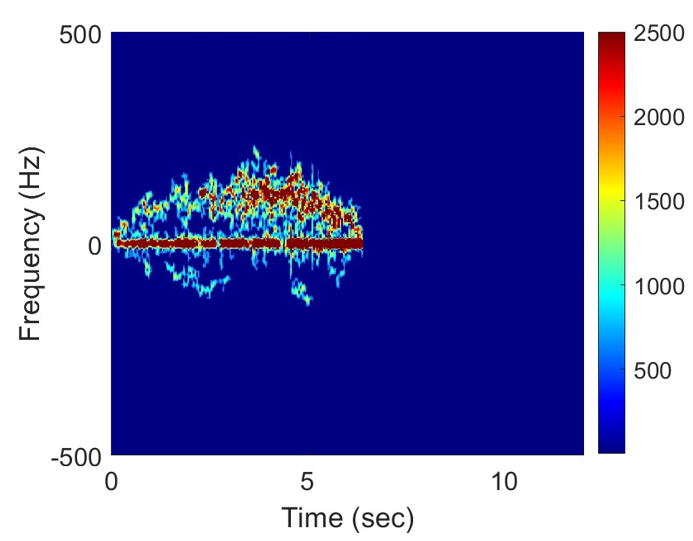}
	 \caption{Pre-Processing Stage Implementation}
              {Left: Envelope Detection, Center: Event Detection, Right: Zero-Padded Image}
    \label{PreProcessing_Image.2}
\end{figure}

Envelope detection is a technique used to extract the envelope of the modulated extended MD  signal, which essentially captures the highest and lowest amplitude variations at all time instances. When a extended MD  image is utilized as the primary signal, the Van Dorp and Groen percentile technique can be employed to reduce the radar return into a single array. This percentile technique calculates the cumulative amplitude distribution for each temporal slice\cite{articleB1}. The common approach to implementing this technique with a extended MD return is to find the upper $u_i$, central $c_i$, and lower $l_i$ envelopes through percentile multiplication with the sum of intensities with time $I(n)$ as follows,

\begin {equation}  \label{a}
u_i (n)=0.97*I(n),
\end {equation}
\begin {equation}  \label{b}
c_i (n)=0.50*I(n),
\end {equation}
and
\begin {equation}  \label{c}
l_i (n)=0.03*I(n).
\end {equation}
Because background noise cannot be completely filtered out of the extended MD  during the smoothing process, shown in Figure \ref{PreProcessing_Image.2} Left, we propose utilizing two central envelopes, starting from opposite ends of the extended MD image, and averaging across them in time to obtain the `real' central envelope. 

After obtaining the central envelope, the STA/LTA ratio is calculated continuously at each time sample $n$ for every $j$th pixel along the central envelope $c_i$ as $R={\frac{STA(n)}{LTA(n)}}$\cite{articleB2}, where
\begin {equation}  \label{d}
STA(n)=\frac{1}{N_1}\sum_{j=n+1}^{n+N_1}{c_i}(j),
\end {equation}
and
\begin {equation}  \label{e}
LTA(n)=\frac{1}{N_2}\sum_{j=n-N_2}^{n}{c_i}(j).
\end {equation}
$N_1$ and $N_2$ are short/long window lengths, respectively. The start time of an event is determined when $STA(n)>\sigma_1$ and $R>\sigma_2$ and ends when $STA(n)<\sigma_3$ and $R<\sigma_2$ where $\sigma_1, \sigma_2, \sigma_3$ are pre-defined detection thresholds. Note that we have considered non-overlapping STA and LTA windows.

Ideally, the triggered event should include all phases, but for the majority of these weak events, the algorithm does not trigger at the beginning because the threshold for ignition of the start time is reached a couple of pixels late. These detection thresholds are set based on the maximum frequency pixel of non-motion observation. To counteract this side effect, we propose an appropriate pre-event time (PEM) selection that ensures the correct start time of the event \cite{articleB3}. This same concept of incorrect ignition due to threshold setting also occurs during detriggering of an event. Because of this, we also set a post-event time (PET) parameter. Optimal PEM and PET durations depend mostly on the application. For the underlying scenario, the extended MD  image spans over twelve-second, PEM=PET=${\frac{Size of Event}{20}}$.

It's worth noting that energy leakage due to  beamforming on radar returns at varying angles can potentially lead to issues with event detection. Figure \ref{PreProcessing_Image.1} demonstrates this by presenting a comparison of zero-padded events. Specifically, the extended MD  image obtained, without beamforming,  struggled to capture the first motion because of the similar intensities across upper and lower spectra, whereas the extended MD  image obtained at $\pm$30\degree angles performed as expected due to sufficient filtering of the other simultaneous motion.


\begin{table}[b]
\centering

\begin{tabular}{ | m{4.7em} | m{3.2em}| m{3.2em} | m{3.2em} | m{3.2em}| m{3.2em} | m{3.2em} | m{3.2em}| m{3.2em} | m{3.2em} |} 
\hline
Predicted v. Actual & Class-1 Bend Down (0\degree) & Class-2 Sit Down (0\degree)  & Class-3 Stand Up (0\degree) & Class-4 Walk Back (0\degree) & Class-5 Walk Back (-30\degree) & Class-6 Walk Back (+30\degree) & Class-7 Walk Forward (0\degree)  & Class-8 Walk Forward (+30\degree) & Class-9 Walk Forward (-30\degree)\\ 
\hline
Class-1 Bend Down (0\degree) & 83.3\% & 4.2\%  & 12.5\%  & 0\%  & 0\%  & 0\%  & 0\%  & 0\%  & 0\% \\ 
\hline
Class-2 Sit Down (0\degree) & 4.2\% & 95.8\% & 0\%  & 0\%  & 0\%  & 0\%  & 0\%  & 0\%  & 0\% \\
\hline
Class-3 Stand Up (0\degree) & 0\% & 0\% & 93.3\%  & 0\%  & 0\%  & 0\%  & 6.7\%  & 0\%  & 0\%  \\
\hline
Class-4 Walk Back (0\degree) & 0\% & 0\% & 0\%  & 100\%  & 0\%  & 0\%  & 0\%  & 0\%  & 0\% \\  
\hline
Class-5 Walk Back (-30\degree) & 0\% & 0\% & 0\%  & 0\%  & 100\%  & 0\%  & 0\%  & 0\%  & 0\% \\
\hline
Class-6 Walk Back (+30\degree) & 0\% & 16.7\% & 0\%  & 0\%  & 0\%  & 83.3\%  & 0\%  & 0\%  & 0\% \\ 
\hline
Class-7 Walk Forward (0\degree) & 0\% & 0\% & 0\%  & 0\%  & 0\%  & 0\%  & 100\%  & 0\%  & 0\% \\
\hline
Class-8 Walk Forward (+30\degree) & 0\% & 0\% & 0\%  & 0\%  & 0\%  & 0\%  & 0\%  & 100\%  & 0\%  \\
\hline
Class-9 Walk Forward (-30\degree) & 0\% & 0\% & 0\%  & 0\%  & 0\%  & 0\%  & 0\%  & 0\%  & 100\%  \\
\hline

\end{tabular} 

\captionof{table}{Confusion Matrix for one and two-person data after image segmentation and zero-padding. The 0\degree, +30\degree, and -30\degree \space indicate the angle at which the activity was captured with respect to the radar broadside}
\label{table:2}
\end{table} 

\section{Micro-Doppler Classification}
\label{Feature extraction and classification}

The CNN is a widely used neural network for image classification due to its ability to automatically select image features through simple convolution and nonlinear activation operations. The CNN's architecture, shown in Fig. \ref{PreProcessing_Image.1}, utilizes three input modalities including the extended MD images collected from three angular locations, namely the array broadside and  $\pm30\degree$ with respect to the array broadside. Each extended MD image is of size $128 \times128 \times 3$ and is passed through a 3-layer CNN consisting of 16 filters in each layer, employing $3\times3$ convolution. It is noted that the CNN receives the MD image after cropping and zero-padding. As mentioned previously, individual motions are cropped from extended MD images using STA/LTA, start/end timestamps. However, the length of the cropped events may vary depending on the duration of the activity. For instance, walking may have a longer duration than bending 
motion. These dimensional mismatches can be problematic for the CNN since it requires all images to be $128 \times128 \times 3$. To address this issue, we zero-pad the cropped images to $600 \times600 \times 3$ and then resize them to $128 \times128 \times 3$. For two-person data, cropping is performed similarly using STA/LTA start/end times to isolate desired motions. 

The outputs of the three networks corresponding to the data inputs are concatenated and fed to a dense layer followed by an output layer of size $9$, which represents the number of possible activities. The output layer is a one-hot encoded vector where the location of a single 1 in the output vector indicates a specific activity. The internal layers of the network use ReLU activation function, while the softmax activation function is used for the output layer.

 \section{Experimental Results} \label{Experimental Results}

In this section, we show that the radar system is capable of separating the  MD spectrograms of one-person and two-people such that the CNN is able to adequately classify each individual activity in a transitional movement. We implemented various algorithms to detect multiple transitional activities happening within one extended MD  image. After pre-processing and spatial filtering of the  extended MD spectrograms, we were able to isolate different activities happening within the frame at a desired angular region.

 To train the CNN, we used 80\% of the samples for each class in the training set, where each sample consists of three $128 \times 128 \times 3$ images (one nominal and two beamformed images). The network weights were optimized using the Adam Optimizer with a learning rate of 0.0001 and 25 epochs. Table \ref{table:2} presents the confusion matrix, which shows the percentage of each class that was predicted versus the actual class based on our labeling. It provides a visual representation of the activities that are being incorrectly classified as other activities. The overall accuracy of the network is observed to be around  95.1\%.




\section{Conclusion}
\label{Conclusion}

In this paper, we introduce a novel approach that utilizes the TF representation of radar returns obtained from multiple azimuth angles. This approach enables us to identify combinations of continuous daily activities that are performed simultaneously at different angles. Our method is effective in detecting individual activities from extended MD images, isolating them, and mapping them to their respective angular locations. Notably, our approach yields promising results, particularly when spectrograms are not entirely separable based on angles. In conclusion, our approach provides an efficient means of distinguishing between different motions performed concurrently and has broad applications in various indoor settings.

\section{ACKNOWLEDGMENT}
\label{Problem Formulation}

The authors would like to thank Jessica Levin, John Barr, Nicholas DePrince, and Alexander Milazzo for their assistance with data collection.

\bibliographystyle{IEEEtran}
\bibliography{references}

\end{document}